\documentclass[aps,prd,preprint,superscriptaddress,tightenlines,nofootinbib]{revtex4}

\usepackage{graphicx}
\usepackage{dcolumn}
\usepackage{bm}

\begin{document}

\preprint{CLNS 06/1953}
\preprint{CLEO 06-03}

\title{Anti-deuteron Production in $\Upsilon(nS)$ Decays 
       and the Nearby Continuum}

\author{D.~M.~Asner}
\author{K.~W.~Edwards}
\affiliation{Carleton University, Ottawa, Ontario, Canada K1S 5B6}
\author{R.~A.~Briere}
\author{J.~Chen}
\author{T.~Ferguson}
\author{G.~Tatishvili}
\author{H.~Vogel}
\author{M.~E.~Watkins}
\affiliation{Carnegie Mellon University, Pittsburgh, Pennsylvania 15213}
\author{J.~L.~Rosner}
\affiliation{Enrico Fermi Institute, University of
Chicago, Chicago, Illinois 60637}
% ...
\collaboration{CLEO Collaboration}
\noaffiliation

\date{December 11, 2006}

\begin{abstract} 

Using CLEO data, we study the production of the anti-deuteron, $\bar d$, 
in $\Upsilon(nS)$ resonance decays and the nearby continuum.  
The branching ratios obtained are 
${\cal B}^{\mathrm{dir}}(\Upsilon(1S) \to \bar{d} X) 
             = (3.36 \pm 0.23 \pm 0.25) \times 10^{-5}$, 
${\cal B}(\Upsilon(1S) \to \bar{d} X) 
             = (2.86 \pm 0.19 \pm 0.21) \times 10^{-5}$, and 
${\cal B}(\Upsilon(2S) \to \bar{d} X) 
             = (3.37 \pm 0.50 \pm 0.25) \times 10^{-5}$, 
where the ``dir'' superscript indicates that decays produced via 
re-annihilation of the $b\bar{b}$ pair to a $\gamma^*$ 
are removed from both the signal and the normalizing number 
of $\Upsilon(1S)$ decays in order to isolate direct decays 
of the $\Upsilon(1S)$ to $ggg,\,gg\gamma$.  
Upper limits at 90\% CL are given for 
${\cal B}(\Upsilon(4S) \to \bar{d} X) < 1.3\times 10^{-5}$, 
and continuum production 
$\sigma(e^+e^- \to \bar{d} X) < 0.031$ pb.  
The $\Upsilon(2S)$ data is also used to extract a limit on 
$\chi_{bJ} \to \bar{d} X$.  
The results indicate enhanced deuteron production in $ggg,\,gg\gamma$ 
hadronization compared to $\gamma^* \to q\bar{q}$.  
Baryon number compensation is also investigated with the large 
$\Upsilon(1S) \to \bar{d} X$ sample.  

\end{abstract}

\maketitle

\section{Introduction}

Anti-deuteron production has been observed in $e^+ e^-$ collisions 
at both the $\Upsilon(1S)$ \cite{Argus} and $Z$ \cite{ALEPH} 
resonances as well as in a variety of other interactions 
\cite{DorfanE735H1}.  The study of anti-deuterons 
rather than deuterons avoids large backgrounds from 
interactions with beam gas and detector material in colliders 
and nuclear breakup in fixed-target and heavy ion collisions.  
Since the various hadronization processes we wish to explore are 
expected to be charge-symmetric, there is no loss of information 
incurred by studying only the experimentally cleaner anti-deuterons.  

Theoretical descriptions of anti-deuteron formation are generally 
based on a coalescence model, according to which an anti-neutron and 
anti-proton nearby to each other in phase-space bind together \cite{Sato}.  
Simple calculations may be based on empirical anti-baryon production 
rates, but subtleties arise.  Nearby in phase space largely 
means nearby in vector momentum since the hadronization occurs 
in a compact region.  But the finite size of this region and the 
presence of short-lived intermediate resonances, such as the $\bar{\Delta}$ 
quartet, lead to questions concerning the necessary degree of 
coherence, which can only be addressed with further assumptions.  
The combined $\Upsilon(1S), \Upsilon(2S)$ result from ARGUS \cite{Argus} 
as well as an upper limit from OPAL at the $Z$ \cite{OPAL} 
were accommodated by Gustafson and Hakkinen \cite{Gusta} 
on the basis of a string model calculation used to supply 
details of the fragmentation process.  
ALEPH \cite{ALEPH} also compared their recent result to this model but 
limited precision and momentum range proscribe any definitive conclusions.  
A more accurate experimental result is desirable to further refine models.  

Practical limitations of particle identification restrict the 
momentum range over which anti-deuterons may be studied.  
However, the lower mass of the $\Upsilon(1S)$ means that a 
larger fraction of the momentum spectrum is accessible compared 
to experiments at the $Z$ pole.  
Also, baryon production in $\Upsilon(1S)$ decays is known to be enhanced 
relative to continuum hadronization \cite{Behrends}.  
The $Z$ pole provides a generous rate enhancement but the hadronization 
proceeds via an initial $q\bar{q}$ pair just as the $e^+e^-$ 
continuum, whereas the $\Upsilon(1S)$ decays primarily via three gluons 
which may be contrasted with nearby continuum $q\bar{q}$ data.  
Glue-rich $\Upsilon(1S)$ decays might also produce exotic multi-quark states, 
beyond $q\bar q$ and $qqq$ \cite{Rosner}.  
As with anti-deuterons, these may form in a similar coalescence process of 
intermediate hadrons or from the primary $ggg,\,gg\gamma$ hadronization 
\cite{Karliner}.  
It is also interesting to search for evidence of anti-deuteron 
production inconsistent with coalescence.  The frequency with 
which baryon number is compensated via two baryons compared to 
a deuteron accompanying the anti-deuteron may prove useful 
in this regard.  

Our key result will be a much-improved determination of the 
rate of anti-deuteron production from $\Upsilon(1S) \to ggg, gg\gamma$ 
hadronization.  The momentum dependence of production may help discriminate 
production models and is also used to estimate production outside 
our experimentally accessible momentum range.  
Given the larger data samples, we do not need to combine 
$\Upsilon(1S)$ and $\Upsilon(2S)$ production as done \cite{Argus} 
previously, but instead use the $\Upsilon(2S)$ data to limit 
$\chi_{bJ}(1P)$ production of anti-deuterons.  
In addition, we obtain a first limit on anti-deuterons from 
the $\Upsilon(4S)$ and an improved limit on continuum production.

\section{Data and Selection Criteria}

We use data collected with the CLEO detector 
at the Cornell Electron Storage Ring, at or near the energies of the 
$\Upsilon(nS)$ resonances, where $n=1,2,4$.  
The analyzed event samples correspond to integrated luminosities of 
1.2 fb$^{-1}$ on the $\Upsilon(1S)$, 0.53 fb$^{-1}$ on the $\Upsilon(2S)$, 
0.48 fb$^{-1}$ on the $\Upsilon(4S)$, and 0.67 fb$^{-1}$ 
of continuum data from just below the $\Upsilon(4S)$.  
The resonance samples contain a total of 
$21.95 \times 10^6$ $\Upsilon(1S)$, 
$ 3.66 \times 10^6$ $\Upsilon(2S)$, and 
$ 0.45 \times 10^6$ $\Upsilon(4S)$ decays.  

Smaller effective cross-sections on the $\Upsilon(2S)$ and 
$\Upsilon(3S)$ and complications from feed-down decrease 
yields and complicate interpretation of these data.  Thus we 
will emphasize the $\Upsilon(1S)$ sample.  
The $\Upsilon(2S)$ sample is used to limit anti-deuteron 
production from $\chi_{bJ}(1P)$ decays by assuming that the 
$ggg,\,gg\gamma$ production from the $\Upsilon(1S)$ and $\Upsilon(2S)$ 
are identical.  
We choose not to analyze an available $\Upsilon(3S)$ sample since 
the statistical error on a branching ratio would be quite large.  
It would also not be possible to separate the contributions from 
$\Upsilon(1S)$, $\Upsilon(2S)$, $\chi_{bJ}(1P)$ and $\chi_{bJ}(2P)$ 
feed-down from the direct $\Upsilon(3S)$ decays in a meaningful way.  

The four innermost portions of the CLEO detector are immersed in a 1.5 T 
solenoidal field. Charged-particle tracking is provided by a four-layer 
double-sided silicon microstrip detector \cite{Hill} and a 47-layer small-cell 
drift chamber with one outer cathode layer \cite{Peterson}.  
The drift chamber also provides particle identification via specific 
ionization ($dE/dx$) measurements.  Surrounding the drift chamber is a LiF-TEA 
Ring-Imaging CHerenkov (RICH) detector \cite{Artuso}, followed by a CsI(Tl) 
calorimeter \cite{Kubota}.  Most critical in the current analysis are 
the drift chamber, which covers $|\cos\theta| < 0.93$, and the 
RICH detector, which covers $|\cos\theta| < 0.80$, 
where $\theta$ is the polar angle with respect to the $e^+e^-$ beams.  

Our anti-deuteron track selection proceeds as follows.  
First, a candidate charged track must be consistent with originating 
from the interaction point.  
The impact parameter with respect to the nominal collision point 
along the beam direction, $\Delta z$, must satisfy $|\Delta z| < 0.05~$m; 
this distribution is dominated by the physical beam bunch length.  
The impact parameter in the $r-\phi$ plane perpendicular to the beam, 
$\Delta r$, is required to satisfy $|\Delta r| < 0.005~$m.  
Since the transverse beam-size is much smaller, the difference here 
is taken with respect to a time-averaged collision point to account 
for accelerator lattice changes and other effects.  The collision point 
can be stable over many days for a fixed lattice.  
The track must be well-measured, based on the reduced $\chi^2$ of the 
track fit and the fraction of traversed drift-chamber layers with good hits.  
Due to difficulties in reconstructing low-momentum tracks (especially 
considering the large energy loss of the softest anti-deuterons 
in material before the drift chamber) 
and the shrinking $dE/dx$ separation between anti-deuterons and other species 
at high momentum, we only consider tracks with momenta between 
0.45 GeV/$c$ and 1.45 GeV/$c$.  
We will later estimate the amount of signal outside this momentum interval.  

The identification of a quality track as an anti-deuteron relies 
on the ionization energy loss measurement in the drift chamber ($dE/dx$).  
To ensure a high-quality $dE/dx$ measurement, 
we only use tracks with at least 10 charge samplings remaining 
after truncation of the highest 20\% and lowest 5\% of the charge 
samples for each track.  
Further, the track angle with respect to the beamline,  $\theta$, 
must satisfy $|\cos\theta| > 0.2$ in order to avoid large gas-gain 
saturation effects present at normal incidence with respect to the 
chamber wires.  This limit was chosen by examining the behavior of the 
large inclusive deuteron sample and observing where the success of 
the corrections applied to compensate for this saturation begin to degrade.  

The $dE/dx$ measurement is converted to a normalized deviation 
\begin{equation}
   \chi_d \equiv \frac{(dE/dx)_{\mathrm measured} 
                     - (dE/dx)_{\mathrm expected,d}}
                      {\sigma_{dE/dx}} 
\end{equation}
with respect to the ionization expected for a real (anti-)deuteron.  
The $dE/dx$ expected mean and resolution ($\sigma$) 
include dependencies on velocity ($\beta\gamma = p/m$), $\cos\theta$, 
and the number of hits used to obtain the measured $dE/dx$.  
We accept a track as a deuteron candidate if $-2 < \chi_d < +3$; 
the asymmetric cut is chosen to reduce background from the large number 
of protons, which appear at lower values of $dE/dx$.  

To suppress $\pi$ and $p$ background, we impose requirements on the number of 
detected Cherenkov photons in the RICH detector.  
Proton suppression is important since they they are nearest to deuterons 
in ionization, while suppression of pions is added since they are so 
numerous; kaon suppression is not employed.  
For a given particle hypothesis, only photons within 
three standard deviations of the expected ring location are counted; 
we require fewer than five photons for the $\pi$ hypothesis 
and fewer than three photons for the $p$ hypothesis.  
For the entire momentum range, pions are well above Cherenkov 
threshold and give more than 10 detected photons on average, 
while protons cross threshold near $p = 0.9$ GeV/$c$ with the 
mean number of photons increasing with increasing momentum.  

\section{Yield Extraction and Backgrounds}

Our signal is typified by a well-reconstructed track coming from the 
interaction point, with $dE/dx$ consistent with an anti-deuteron.  
We choose to use the distribution of the normalized deviation, $\chi_d$, 
to determine our raw signal yield.  We do this in five 200 MeV/$c$ momentum 
bins spanning $0.45 - 1.45$ GeV/$c$.   

The backgrounds to our anti-deuteron signal are from three main sources.  
The first is particle mis-identification.  For most of the momentum range 
considered, $dE/dx$ separation is good; however, since anti-deuterons 
are very rare compared to the other hadrons, even a small resolution or 
mis-measurement tail may be troublesome.  
Second, spurious hadrons are produced via interactions of beam particles 
or genuine decay products with residual gas in the beampipe or the beampipe 
and inner detector material.  
In practice, this is a much larger issue for deuterons than anti-deuterons 
since the gas and material are matter and not anti-matter.  
This is the primary reason we focus on anti-deuterons in this study.  
Finally, for resonance decays, there is a possible non-resonant contribution 
from the continuum events underlying the $\Upsilon$ resonance peaks.  

\begin{figure}
\includegraphics*[width=3.75in]{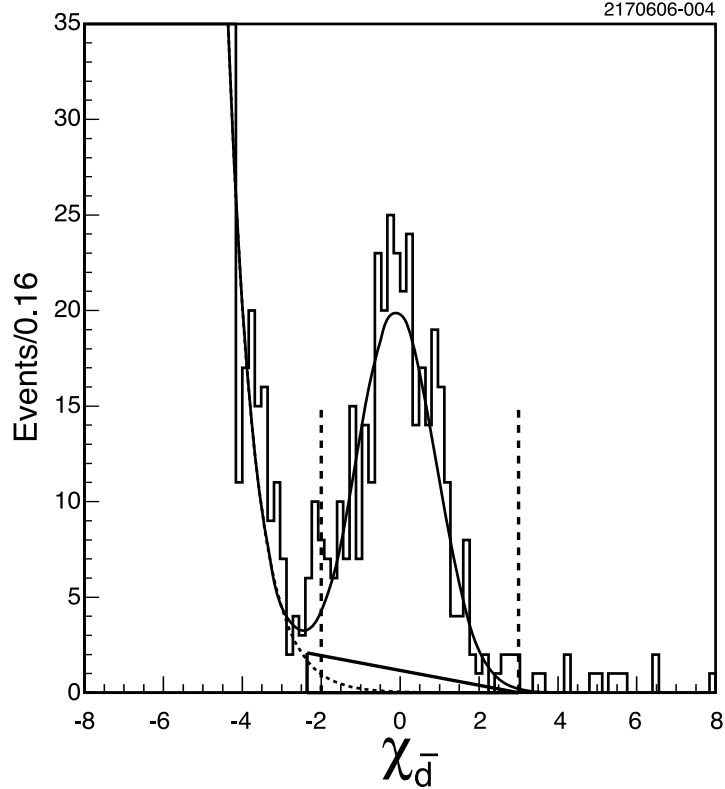}
\caption{The $\chi_d$ distribution in the momentum range $0.45 - 1.45$ GeV/$c$ 
         fit to a Gaussian anti-deuteron signal and a falling background 
         function.  
         The background estimation employs the area of the 
         triangular region between the dashed cuts, 
         as detailed in the text.}
\label{fig:tberr}
\end{figure}

Our raw yield and the particle mis-identification background are determined 
as follows.  We count the total number of entries between $-2 < \chi_d < +3$, 
denoting this as $N$.  To estimate the background from misidentification, 
we fit the $\chi_d$ distribution to a Gaussian signal peak plus 
an exponential background shape, as shown in Fig.~\ref{fig:tberr}.    
The mean and width of the Gaussian are fixed from fits to the larger 
deuteron sample in the data.  
We then define a triangular background shape as shown 
in Fig.~\ref{fig:tberr}.    
The apex lies on the background curve at the $\chi_d$ value 
corresponding to the minimum of the total fit function, 
between the rapidly falling background and the signal peak.  
The triangle is drawn to decrease to zero height at $\chi_d = +3$.  
We denote the area of this triangle between $-2 < \chi_d < +3$ as $A$.  
We then take the central value of the raw yield as $N - A/2$ 
with an error of $\pm A/4$.  
The rapidly falling fit would argue for a lower background, while a small 
accumulation of events at $\chi_d > +3$ balances this.  
In fact, we know little about the background shape other than 
naively expecting a falling shape.  Our method spans the range from $0$ 
to $A$ for the background size within $\pm 2 \sigma$, where $\sigma = A/4$.  

For beam particles or decay products interacting with residual gas 
or beampipe and detector material, the tracks do not peak in both $r-\phi$ 
and $z$ impact parameters as the signal does.  We therefore estimate 
the underlying background from these sources by looking 
at the impact-parameter sidebands.  
These backgrounds are small and assumed to be flat for purposes 
of the modest extrapolation underneath the peaks.  

The non-resonant contribution is estimated through analysis of our 
off-resonance data sample.  Since this contribution is found to be small, 
we use data taken below the $\Upsilon(4S)$ resonance to subtract 
it for all $\Upsilon$ resonances.  
The off-resonance data is scaled by a factor accounting for the ratio 
of luminosities and the $1/s$ dependence of the cross section 
with respect to the resonance sample.  
We will also consider the process $\Upsilon(nS) \to \gamma^* \to q\bar{q}$ 
as background, because the physics is the same as the continuum 
$e^+ e^-  \to q\bar{q}$ process, and we are interested primarily 
in the $ggg$ and $gg\gamma$ decays of the $\Upsilon$.  
Thus, when subtracting continuum yields, we adjust the scale 
factor to account for the additional continuum-like events produced 
via re-annihilation via a $\gamma^*$.  
Similarly, the most useful branching ratio will be normalized not to 
the total number of decays, $N(\Upsilon(nS))$, but instead to 
\begin{equation}
  N_{\mathrm{dir}} = N(\Upsilon(nS))
                     (1 - (3+R_{\mathrm had}){\cal B}_{\mu\mu}), 
\end{equation}
counting only the decays proceeding via $ggg,\,gg\gamma$ hadronization 
by excluding dilepton decays, which proceed via a $\gamma^*$ as well.  
Here, ${\cal B}_{\mu\mu} = {\cal B}(\Upsilon(nS) \to \mu^+\mu^-)$ 
and $R_{had} = \sigma(e^+ e^- \to\;{\rm hadrons})
              /\sigma(e^+ e^- \to \mu^+\mu^-)$.  
We use $R_{had} =3.56 \pm 0.01 \pm 0.07 $ \cite{Ammar} 
and ${\cal B}_{\mu\mu} = (2.49 \pm0.02 \pm 0.07)\%$ \cite{Danko}.

We refer to the resulting branching ratio as the ``direct'' one.  
This branching ratio is equivalent to 
\begin{equation}
{\cal B}^{\mathrm{dir}}(\Upsilon(1S) \to \bar{d} X) = 
\frac {{\cal B}(\Upsilon(1S) \to ggg,\,gg\gamma \to \bar{d} X)}
      {{\cal B}(\Upsilon(1S) \to ggg,\,gg\gamma \to         X)}.
\end{equation}
This will be our central result concerning $ggg,\,gg\gamma$ hadronization.  
For completeness, we also present a conventional branching ratio without 
the modifications described in the preceding paragraph.  
Since we will find that continuum production of anti-deuterons is small, 
while non-$ggg,\,gg\gamma$ decays of the $\Upsilon(1S)$ are significant, 
the direct branching ratio will be larger than the conventional 
inclusive one.  

Results of the above yield and background determinations are summarized 
in Table~\ref{tab:dbarydmom}.  

\section{Detection Efficiency}

\subsection{Tracking and RICH Efficiency}

We use Monte-Carlo event samples to study the anti-deuteron efficiency 
of our tracking and RICH criteria.   
We cannot use anti-deuterons in our simulations since this particle 
is not included in GEANT, which is the basis of CLEO Monte-Carlo software.  
However, we do not expect significant differences between deuteron 
and anti-deuteron behavior since both the RICH detector and the tracking 
are largely charge-independent, as are our selection criteria.  
Nuclear interactions do distinguish $d$ and $\bar{d}$, but given our 
large final errors, we may safely neglect this effect as well.  
In particular, annihilations in the beampipe or silicon vertex detector 
are estimated to be negligible given our statistics.  
We also note that our $\Upsilon$ decay hadronization models produce 
very few deuterons; this leads us to choose the following techniques.  

Our first Monte-Carlo sample consists of events with one deuteron 
and no other detector activity.  The second consists of 
overlaying the preceding type of ``single-track'' events on top 
of a real $\Upsilon(1S)$ decay from data.  The former likely overestimates 
the efficiency due to the quiet detector environment, while the latter 
likely underestimates it due to excess activity (since nothing is removed 
from the full decay when the signal track is added in).  
We obtain tracking efficiencies of about 70\%, with a 10\% relative 
difference between the two methods.  

We average the efficiencies of our two Monte-Carlo samples, taking 
one-quarter of the difference between them as a systematic uncertainty.  
The resulting efficiency is fairly flat, except in the lowest momentum bin 
of $0.45 - 0.65$ GeV/$c$, where it decreases by about 10\% of itself.  
By re-weighting Monte-Carlo events according to the momentum distribution 
of the data across this bin, we find that we are not very sensitive 
to the detailed spectrum, but we do add an additional systematic error 
for this effect.  

Finally, our signal is consistent with being flat vs.~$\cos\theta$ 
in the accepted range $0.20 < |\cos\theta| < 0.93$; we assume it is 
flat when evaluating the effect of our fiducial cut on the track-finding 
efficiency.  

\subsection{$dE/dx$ Efficiency}

The CLEO Monte-Carlo simulation of $dE/dx$ measurements is done 
at the track-level and is based on the calibrated expected means 
and resolutions.  However, (anti-)deuterons have not been searched 
for in any other CLEO analysis to date.  
Since the calibration is quite challenging for the very high ionization 
of the lower momentum anti-deuterons, the $dE/dx$ calibration was redone 
for the data samples used here.   These new calibrations offer less bias 
versus parameters such as angle and momentum than the standard versions.  
But, as a result, the CLEO $dE/dx$ simulation designed for the standard 
calibration is not well-suited for our analysis.  
Instead, we use deuterons from real data, produced by beam-gas interactions, 
to estimate the $dE/dx$ efficiency.  Our impact parameter cuts ensure that 
these tracks are geometrically similar to signal tracks.  

\begin{figure}
\includegraphics*[width=3.75in]{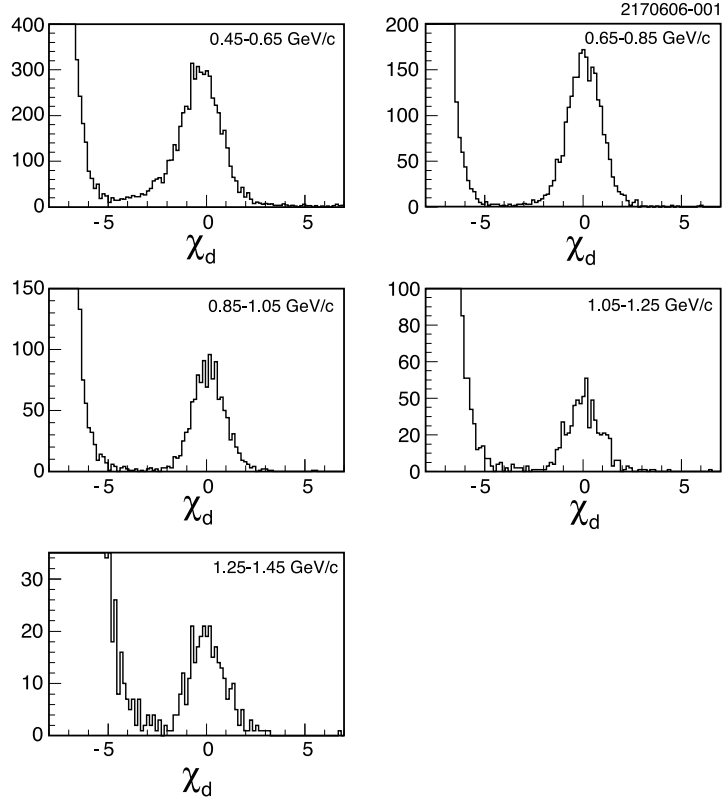}
\caption{Deuteron $\chi_d$ distributions ($dE/dx$ normalized deviation) 
         in different deuteron momentum bins.}
\label{fig:ddedxshapeeff}
\end{figure}

Fig.~\ref{fig:ddedxshapeeff} shows the deuteron $\chi_d$ distributions 
for all five momentum bins; these are mostly deuterons from beam-gas 
interactions or nuclear interactions in the detector.  
We define the $dE/dx$ signal efficiency as 
$\epsilon_{dE/dx} = N_{\mathrm sig}/{N_d}$, 
where $N_{\mathrm sig}$ is the yield in the interval $-2 < \chi_d < +3$ and 
$N_d$ is our estimate of the total number of deuterons for all $\chi_d$.  
We estimate $N_d = (N_{\mathrm tot} - N_{\mathrm tail}/2) \pm 
N_{\mathrm tail}/4$.  
Here $N_{\mathrm tot}$ is the yield in the interval $-5 < \chi_d < +5$ 
for the lowest two bins, in $-4 < \chi_d < +4$ for the next 
two, and in $-3 < \chi_d < +4$ for the highest momentum bin.  
$N_{\mathrm tail}$ is designed to include a possible tail from the 
large background at low $\chi_d$ and is taken as the yield in 
the following momentum-dependent intervals: 
$-5 < \chi_d < -4$ for the first two momentum bins, 
$-4 < \chi_d < -3$ for the next two momentum bins, and 
$-3 < \chi_d < -2$ for the last bin.  
The efficiencies are all about 97\%, except for the lowest 
momentum bin, where is it about 88\% due to the low-side 
resolution tail.  
Our systematic error on the $dE/dx$ efficiency comes from propagating 
the error on $N_{d}$ quoted above.  

The width of $\chi_d$ varies with momentum even after our 
re-calibration, especially in the lowest momentum bin.  
We determine our sensitivity by re-weighting the momentum 
distribution in this bin to better reflect the data, and 
include an additional systematic error on the efficiency.  

\subsection{Systematic Uncertainty Summary}

We now summarize our systematic uncertainties; 
in each case, we give the range across the five 
momentum bins.  
The total efficiency uncertainty, including track-finding, 
selection criteria, and yield extraction from the $\chi_d$ plot, 
ranges from 4.5\% to 16\%.  In addition to systematic issues 
discussed earlier in this Section, we also considered the agreement 
between data and MC simulations of the number of photons associated 
with tracks in the RICH detector and the stability of results for 
variations in track-selection criteria.  
The number of $\Upsilon(1S)$ [$\Upsilon(2S)$] in our data sample has an 
uncertainty of 1.4\% [1.5\%] and the continuum luminosity is known to 2\%.  
The resulting total systematic uncertainties range from 6.1\% to 16.0\%.  
These are on average comparable to the statistical uncertainties 
in the case of the $\Upsilon(1S)$ result, 
and smaller that statistical errors in all other cases.

\section{Results}

\subsection{Anti-deuteron Production in $\Upsilon(1S)$}

Table~\ref{tab:dbarydmom} shows the observed number of events from 
$\Upsilon(1S)$ resonance data, off-resonance data, and $\Delta r$ 
and $\Delta z$ sidebands in the on-resonance data.  
After subtracting the latter sidebands and properly-scaled continuum 
contributions, and correcting for efficiency, we get the number of 
$\bar d$ events produced by $\Upsilon(1S)$ decays shown in 
Table~\ref{tab:dbdp}. 
The direct yield column includes a larger continuum scale factor which 
accounts for the contribution in which $b\bar b$ re-annihilate 
to a virtual photon and form a $q\bar{q}$ pair whose fragmentation products 
contain an anti-deuteron.  
We use this column to get the yield from so-called {\it direct} 
decays mediated by $ggg$ and $gg\gamma$ hadronization.  

\begin{table}[hbtp]
\centering \caption{Anti-deuteron yields and backgrounds for 
                     $\Upsilon(1S)$ data in momentum bins.} 
\vspace{0.3cm}
\begin{tabular}{| c | l | l | l | l |}
\hline
Momentum (GeV/$c$) & On $\Upsilon(1S)$ & Continuum & $\Delta r$ sideband & $\Delta z$ sideband \\
\hline
$0.45 - 0.65$  & $60.4 \pm 7.9$  &  $2.0 \pm 1.4$       &  $9.0 \pm 3.0$  
  &  $6.0 \pm 2.5$ \\
$0.65 - 0.85$  & $77.9 \pm 9.2$  &  $1.0 \pm 1.0$       &  $2.0 \pm 1.4$  
  &  $4.0 \pm 2.0$ \\
$0.85 - 1.05$  & $71.0 \pm 8.9$  &  $0.0^{+1.2}_{-0.0}$ &  $1.0 \pm 1.0$  
  &  $0.0^{+1.2}_{-0.0}$ \\
$1.05 - 1.25$  & $58.1 \pm 7.8$  &  $0.0^{+1.2}_{-0.0}$ &  $0.0^{+1.2}_{-0.0}$  
  &  $0.0^{+1.2}_{-0.0}$ \\
$1.25 - 1.45$  & $46.4 \pm 7.2$  &  $3.0 \pm 1.7$       &  $0.0^{+1.2}_{-0.0}$  
  &  $2.0 \pm 1.4$ \\
\hline
\end{tabular}
\label{tab:dbarydmom}
\end{table}

\begin{table}[hbtp]
\centering \caption{Efficiency-corrected anti-deuteron yields and 
                    differential branching rations for $\Upsilon(1S)$ data 
                    in momentum bins.} 
\vspace{0.3cm}
\begin{tabular}{| c | l | l | l |}
\hline
Momentum (GeV/$c$) & Corr'd Yield & Corr'd Direct Yield & $d{\cal B}^{\mathrm{dir}}/dp$  ($10^{-5}$ $c$/GeV)\\
\hline
$0.45 - 0.65$  & $ 85.6 \pm 14.3$  &  $ 82.1 \pm 16.4$ 
               & $  2.2 \pm  0.5 \pm 0.2$\\
$0.65 - 0.85$  & $111.3 \pm 14.2$  &  $109.7 \pm 15.0$ 
               & $  3.0 \pm  0.4 \pm 0.2$\\
$0.85 - 1.05$  & $106.2 \pm 13.8$  &  $106.2 \pm 14.5$ 
               & $  2.9 \pm  0.4 \pm 0.3$\\
$1.05 - 1.25$  & $ 92.5 \pm 12.6$  &  $ 92.5 \pm 13.8$ 
               & $  2.5 \pm  0.4 \pm 0.2$\\
$1.25 - 1.45$  & $ 60.2 \pm 12.7$  &  $ 55.5 \pm 14.1$ 
               & $  1.5 \pm  0.4 \pm 0.2$\\
\hline
\end{tabular}
\label{tab:dbdp}
\end{table}

\begin{figure}
\includegraphics*[width=3.75in]{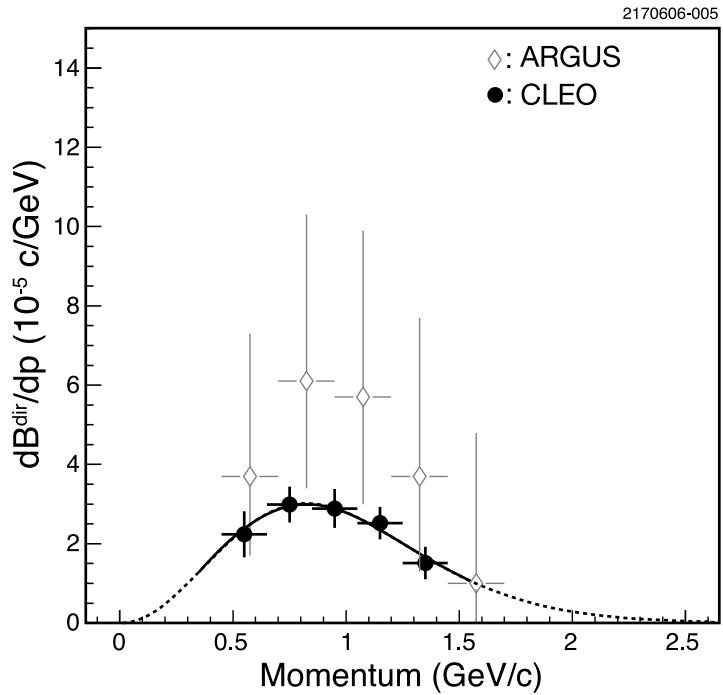}
\caption{Momentum dependence of anti-deuteron production in $\Upsilon(1S)$ 
         decays observed by CLEO (filled circles) and ARGUS (open diamonds).  
         The solid line shows the fit described in the text; the 
         dashed portion is an extrapolation beyond the momentum range 
         which we observe.}
\label{fig:1sfireball}
\end{figure}

To get the anti-deuteron yield in the full momentum range, we fit to the 
Maxwell distribution as used in fire-ball models \cite{Hagedorn}, 
\begin{equation}
 f(p)\equiv a \beta^2 \exp(-E/b),
\end{equation}
where $\beta = pc/E$, and $a$ and $b$ are free parameters.  
We include as a systematic uncertainty the effect of variations of 
the shape parameter $b$ within the statistical errors of the fit; 
we do not include any systematic uncertainty for the accuracy 
of the model itself.  
The resulting fit to the CLEO data is shown, along with the earlier ARGUS 
result, in Fig.~\ref{fig:1sfireball}.  Much of the CLEO systematic 
error is correlated point-to-point, but statistics still dominate
the uncertainty.  Note that the ARGUS data extends to higher momentum 
due to their time-of-flight system for particle identification.  
We also note that ARGUS combined $\Upsilon(1S)$ and $\Upsilon(2S)$ 
yields to extract a more precise $ggg$ rate; it is not clear 
what was assumed about possible $\chi_{bJ}$ production.  
However, it seems most likely that the difference in our 
results is largely statistical; ARGUS has 19 signal events 
from both resonances combined.  

The final branching ratio per direct $\Upsilon(1S) \to ggg, \; gg\gamma$ 
decay is 
\begin{equation}
 {\cal B}^{\mathrm{dir}}(\Upsilon(1S) \to \bar{d} X) 
         = (3.36\pm0.23\pm0.25) \times 10^{-5}.   
\end{equation}
For this calculation, we have used only the number of $\Upsilon(1S)$ 
which decay via $ggg,\,gg\gamma$ as our normalization and subtracted 
a small amount of yield due to $b\bar{b}$ re-annihilation to $\gamma^*$ 
based on the observed off-resonance continuum yield.  
The more inclusive ``conventional'' branching ratio result is 
\begin{equation}
 {\cal B}(\Upsilon(1S) \to \bar{d} X) 
         = (2.86\pm0.19\pm0.21) \times 10^{-5}.   
\end{equation}

\subsection{$d$ Production in $\Upsilon(1S)$}

Given that the deuteron signal is expected to be identical 
to the anti-deuteron signal, but with very much larger backgrounds, 
analyzing for deuterons would not contribute much statistically 
to this analysis.  
However, we can use deuteron production as a consistency check on our 
anti-deuteron measurement.  
We make this comparison for the restricted momentum range $0.6-1.1$ GeV/$c$ 
where the signal-to-noise is best.

\begin{figure}
\includegraphics*[width=3.75in]{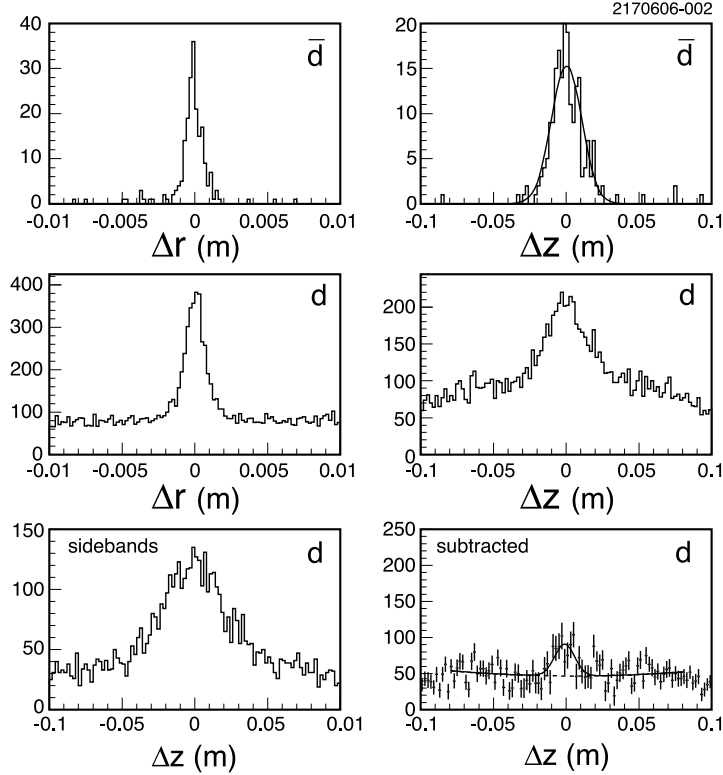}
\caption{Anti-deuteron and deuteron sample distributions for the 
         momentum range $0.6-1.1$ GeV/$c$ in the $\Upsilon(1S)$ data.  
         Top row: $\Delta r$ and $\Delta z$ for anti-deuterons.  
         Middle row: $\Delta r$ and $\Delta z$ for deuterons.  
         Bottom left: $\Delta z$ for deuterons in $\Delta r$ sidebands.  
         Bottom right: Fit to $\Delta z$ for deuterons after subtraction 
         of the $\Delta r$ sidebands in the previous panel.  
         The signal Gaussian width is fixed to the anti-deuteron 
         signal width from the fit to data in the upper right panel.  
         }
\label{fig:dyieldmethod1bin6}
\end{figure}

Since none of the backgrounds described above peak in both 
$\Delta r$ and $\Delta z$, we use a sideband subtraction to 
remove them.  Empirically, we observe that $\Delta r$ sidebands are 
very flat and we therefore subtract $\Delta r$ sidebands from the 
good $\Delta r$ sample and fit the resulting $\Delta z$ distribution 
to a Gaussian peak plus a polynomial background.  
This procedure is displayed in Fig.~\ref{fig:dyieldmethod1bin6}.  
Here, we only use deuterons which satisfy  $-2 < \chi_d < +3$ 
and ignore the small backgrounds from other particle types.  
The resulting deuteron yield is $352.8 \pm 88.6,$ 
about 1.7 standard deviations from the anti-deuteron yield 
of $201.0 \pm 14.2$ in the same 0.6 - 1.1 GeV/$c$ momentum range.  
The $\Delta z$ width is somewhat narrower but similar to that 
for anti-deuterons.

\subsection{Discussion of $\bar d$ Baryon Number Compensation} 

\begin{figure}
\includegraphics*[width=3.75in]{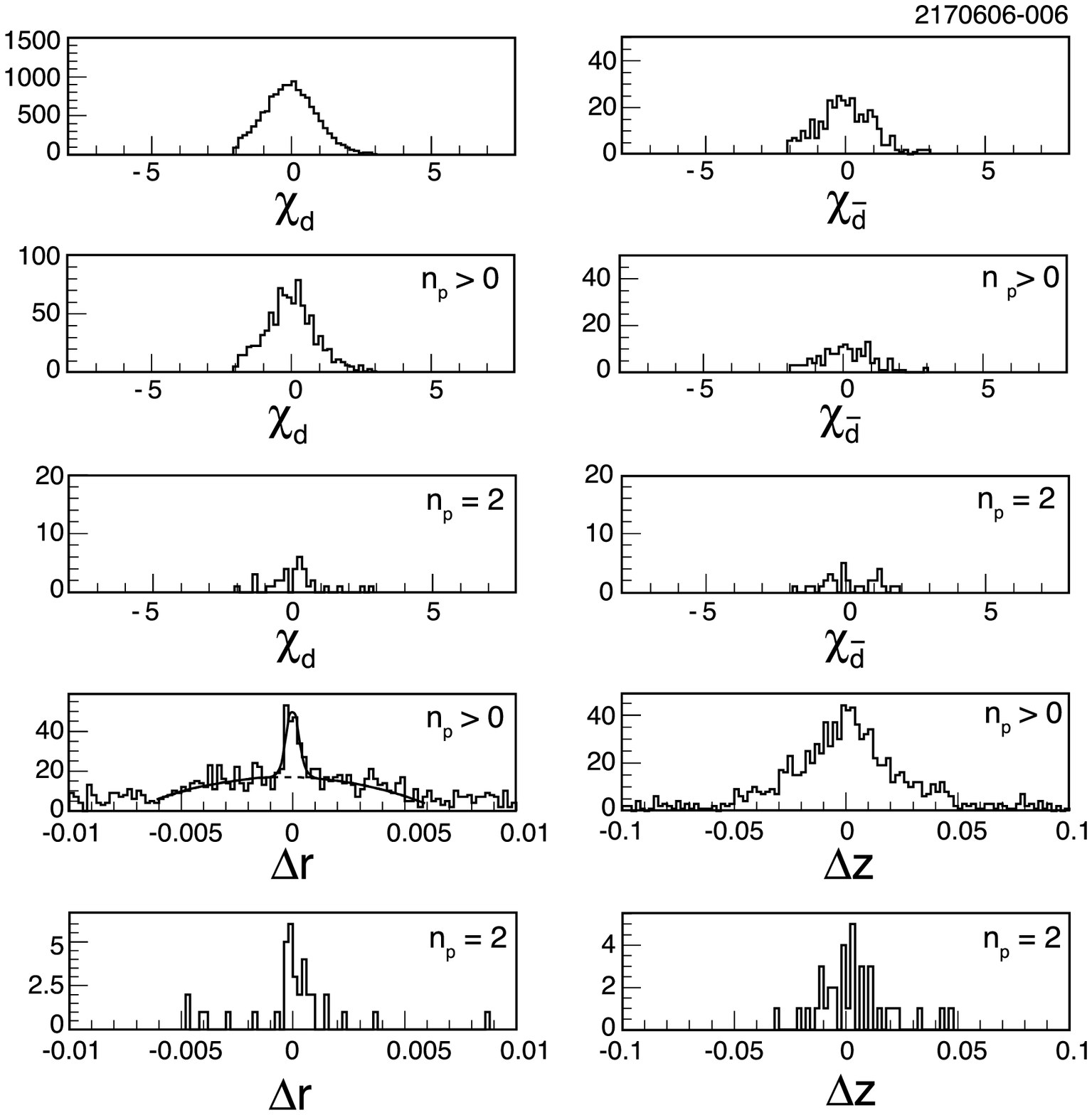}
\caption{$d$ and $\bar d$ yields in the momentum range 0.45-1.45 GeV/c 
         with additional criteria as described.  
   Top row: $\chi_d$ for deuterons (left) and anti-deuterons (right) 
     with all standard cuts, including $-2 <\chi_d < +3$.  
   Second row: As above, but requiring at least one anti-proton (left) 
     or proton (right) candidate in addition.  
   Third row: As above, but requiring two anti-proton (left) 
     or two proton (right) candidates instead.  
   Fourth row: $\Delta r$ and $\Delta z$  for deuterons with $n_p>0$.  
   The curve on the left shows a fit to a signal Gaussian 
     plus a polynomial background.  
   Fifth row:  $\Delta r$ and $\Delta z$  for deuterons with $n_p=2$.  
   }
\label{fig:dsigmanpbargt0}
\end{figure}

Another way to explore consistency of the $d$ and $\bar d$ yields 
is to employ baryon number conservation.  
Assuming many of the $\bar d$ ($d$) events are compensated by 
$pp$ or $pn$ ($\bar{p}\bar{p}$ or $\bar{p}\bar{n}$), 
requiring at least one $p$ ($\bar{p}$) in the event may 
decrease background appreciably.  
As shown in Fig.~\ref{fig:dsigmanpbargt0} and summarized in 
Table~\ref{tab:protoncuts}, after the standard 
selection criteria, we begin with 13140 deuterons and 
338 anti-deuterons candidates (signal plus background).  
Our proton identification requirements are: 
$|\chi_p| < 4$ for $0.30 - 0.85$ GeV/$c$, and 
$|\chi_p| < 3$ for $0.85 - 1.15$ GeV/$c$.  
Here, $\chi_p$ is a normalized $dE/dx$ deviation with respect to the 
proton hypothesis and we do not accept candidates outside the 
two contiguous momentum windows indicated above.  
With this definition of protons, we can study the effect of cuts on 
the number, $n_p$, of protons (anti-protons) in anti-deuteron (deuteron) 
events.  If we require $n_p >0$, 
we are left with 149 $\bar d$ and 898 $d$; 
while the non-decay deuterons are very much reduced, the asymmetry 
indicates residual background from random anti-protons not associated 
with any baryon number compensation.  
This is further verified by examining the $\Delta r$, $\Delta z$ 
distributions in the fourth row of Fig.~\ref{fig:dsigmanpbargt0}.  
The excess is consistent with a spurious deuteron in coincidence with 
a real physics event containing the anti-proton.  
A fit to the $\Delta r$ distribution yields $122.8 \pm16.9$ events 
for the sharp peak, which is now consistent with the anti-deuteron yield.  
Adding a requirement of $n_p \ge 2$, we obtain 31 $\bar d$ and 35 $d$, 
which are quite consistent with equality, implying that most spurious 
$d$ have been removed.  (Note that we never observed $n_p > 2$).  
The remaining $d$ events peak well both in $\Delta r$ and $\Delta z$, 
as expected.  

\begin{table}[hbtp]
\centering \caption{Anti-deuteron and deuteron yields from $\Upsilon(1S)$ data 
                    with requirements on accompanying 
                    protons and anti-protons.} 
\vspace{0.3cm}
\begin{tabular}{|c|c|c|}
\hline
Standard Cuts, plus:  & \# anti-deuteron candidates & \# deuteron candidates \\
\hline
     ---  & 338 & 13140 \\
 $>0$ $p$ / $>0 \bar{p}$   & 149 &   898 \\
 $\ge2 p$ / $\ge2 \bar{p}$ &  31 &    35 \\
\hline
\end{tabular}
\label{tab:protoncuts}
\end{table}

Returning to just the clean anti-deuteron sample, we can take a 
more quantitative look at baryon number conservation.  
By considering the effect of proton-finding efficiency we 
can determine approximately how baryon compensation is distributed among 
$pp$, $pn$, $np$, $nn$; we will consider compensation by a $d$ later.  

In the $-2 < \chi_d < +3$ anti-deuteron signal region, 
we observe 338 candidate events from the $\Upsilon(1S)$ data sample.
Each of these events contains only 1 anti-deuteron.  
Among these 338 events: 189 events contain no protons,  
118 events contain 1 proton, and 31 events contain 2 protons.  
Assuming $\bar d$ is compensated by $pp$, $pn$, $np$, $nn$ (neglecting $d$ 
for now) with an equal probability of $25 \%$, we may estimate what is 
expected, given a proton-finding efficiency.  
We cannot distinguish $pn$ and $np$, but it makes the assumed equality 
clearer to list them separately.  
For the proton identification cuts given above, the efficiency is about 60\%, 
where we assume the spectrum of protons accompanying deuterons is similar 
to the inclusive proton spectrum.  Part of the loss of efficiency is due 
to protons with momenta outside our accepted range of $0.30 - 1.15$ GeV/$c$.   

Folding in this approximate efficiency, we predict 30 events containing 
2 protons, with 31 observed, and 142 events containing 1 proton, 
with 118 observed, as summarized in Table~\ref{tab:protonpred}.  
Within the limits of our uncertainties and assumptions, 
our data is consistent with baryon number conservation occurring 
with roughly equal probabilities for accompanying $pp$, $pn$, $np$, or $nn$.  

\begin{table}[hbtp]
\centering \caption{Anti-deuteron yields from $\Upsilon(1S)$ decay  
                    for various numbers of accompanying protons.}
\vspace{0.3cm}
\begin{tabular}{|l|c|c|c|c|}
\hline
          &  All & $\# p = 0$ &  $\# p = 1$ &  $\# p = 2$ \\ 
\hline
Observed  &  338 & 189 & 118 & 31 \\
\hline
Predicted &  --- & 166 & 142 & 30 \\
\hline
\end{tabular}
\label{tab:protonpred}
\end{table}

\begin{figure}
\includegraphics*[width=3.75in]{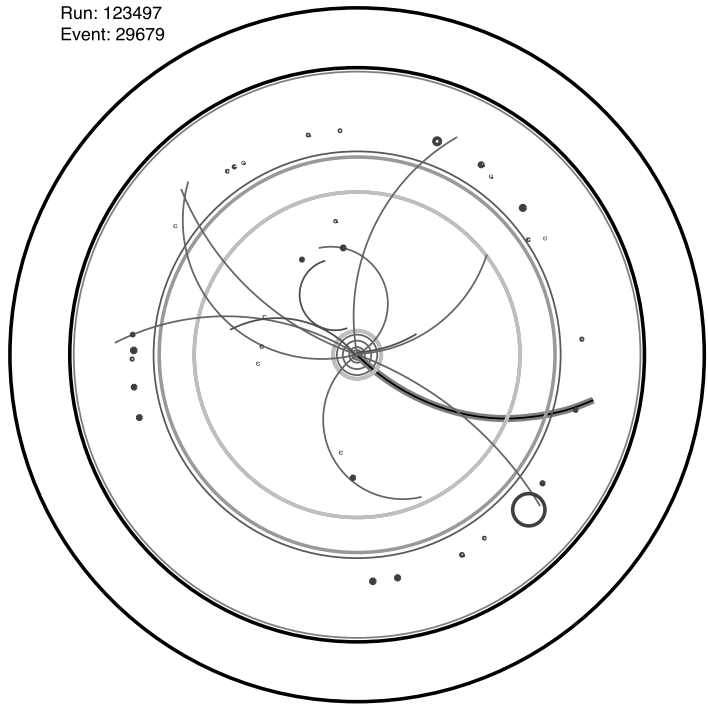}
\caption{An $\Upsilon(1S) \to 3\pi^+ 3\pi^- d\bar d X$ candidate 
         event in the CLEO detector, viewed along the beam axis.  
         The $d$ track ($p = 0.55$ GeV/$c$) is highlighted.   
         The $\bar d$ track ($p = 0.84$ GeV/$c$) is the one with most 
         energetic calorimeter shower; 
         the circle size is proportional to the shower energy.  
         The difference between the center-of-mass energy 
         and the total energy of observed particles, based 
         on tracking and particle identification, 
         is about 120 MeV.  }
\label{fig:29679}
\end{figure}

It is also interesting to look for compensation of an anti-deuteron 
by a deuteron; this is found to occur at the 1\% level.  
Fig.~\ref{fig:29679} shows one of our four possible $d\bar d$X events, 
which is nearly fully reconstructed.  
Inspection of these four $d\bar{d}$ candidate events reveals that one 
of them is consistent with a through-going deuteron track (presumably 
from a cosmic-ray interaction) faking a 
$d\bar{d}$ pair.  The remaining three are consistent with true $d\bar{d}$.  
Through-going deuterons might constitute a non-negligible 
background to our anti-deuteron yield, if the inward deuteron track 
passed our anti-deuteron cuts, but the outgoing deuteron fails.  
Therefore, we have searched for such events with relaxed cuts.  
We find none in the $\Delta r$ and $\Delta z$ sidebands, 
nor do we find any anti-deuteron candidate events where there 
is a lower-quality track candidate failing our cuts back-to-back 
with our candidate track.  
We conclude that this faking mechanism is rare and the one event 
seen was a somewhat unlikely occurrence for our data sample size.

\subsection{$\bar d$ Production in 2S, 4S and Continuum}

We now summarize results from other $\Upsilon$ resonances and 
the continuum.  

In $\Upsilon(2S)$ data, 69 anti-deuteron events are observed.  
This sample has the same background sources as the 
$\Upsilon(1S)$ data, but contains several possible sources 
of anti-deuteron signal.  These include: 
(i) $\Upsilon(2S) \to \Upsilon(1S)\;X$, followed by 
$\Upsilon(1S) \to \bar{d}\,X'$, 
(ii) $\Upsilon(2S) \to ggg,\,gg\gamma$, and 
(iii) $\Upsilon(2S) \to \gamma\,\chi_{bJ}\,X$, followed by 
$\chi_{bJ} \to \bar{d}\,X'$.  
We may subtract process (i) based on known branching ratios.  
Separating (ii) and (iii), for example by looking for the transition 
$\gamma$ in (iii), is not feasible with our limited statistics.  
However, we can {\it assume} that the rate for the direct decay (ii) 
is equal to the analogous $\Upsilon(1S)$ process, and look for 
any excess from $\chi_{cJ}$ decays.  This is interesting since the 
$\chi_{cJ}$ decay via $gg$ for $J=0,2$ and via $gq\bar{q}$ for $J=1$ 
and thus access distinct hadronization processes.  

After background subtractions analogous to the $\Upsilon(1S)$ case, 
we find $58.3 \pm 8.6$ signal events, which translates to
\begin{equation}
  {\cal B}(\Upsilon(2S)\to \bar d +X) 
    = (3.37 \pm 0.50 \pm 0.25)\times10^{-5}.  
\end{equation}
To isolate this rate, we subtract contributions from the processes 
$e^+e^-\to\Upsilon(2S)\to \pi\pi\Upsilon(1S)$ and 
$e^+e^-\to\Upsilon(2S)\to \gamma\gamma\Upsilon(1S)$ 
(two-photon transitions via the $\chi_{bJ}$ states) assuming 
that these processes dominate inclusive $\Upsilon(1S)$ production.  
We must further assume that direct $ggg,\,gg\gamma$ decays of the 
$\Upsilon(2S)$ produce antideuterons at the same rate as 
the $\Upsilon(1S)$.   We are left with an insignificant excess, 
and extract a 90\% CL upper limit for a weighted average 
of the $\chi_{bJ}$ states of 
\begin{equation}
  \sum_J ( {\cal B}(\Upsilon(2S) \to \gamma \chi_{bJ}(1P)) \times 
           {\cal B}(\gamma\chi_{bJ}(1P) \to \bar{d} X) ) / 
  \sum_J   {\cal B}(\Upsilon(2S) \to \gamma \chi_{bJ}(1P)) 
           < 1.1 \times 10^{-4}.  
\end{equation}
This limit is not stringent enough to draw firm conclusions 
on anti-deuteron production in these distinct $gg$ and $gq\bar{q}$ 
hadronization processes in contrast to $ggg,\,gg\gamma$.  

In $\Upsilon(4S)$ data, 3 $\bar{d}$ candidates are observed.  
Based on $\Delta r$ and $\Delta z$ sidebands and the continuum 
data, we expect 5.2 background events.  
For both this limit and the following continuum production limit, 
we ignore any possible backgrounds in the $\chi_d$ distribution.  
We obtain a 90\% CL upper limit, 
using the Feldman-Cousins method \cite{FC}, of 
\begin{equation}
 {\cal B}(\Upsilon(4S) \to \bar{d} X) < 1.3\times 10^{-5}.  
\end{equation}
This limit is not very stringent in view of the dominance of 
$B\bar{B}$ decays of the $\Upsilon(4S)$.  

A 90\% CL upper limit result for continuum production is also obtained, 
based on 6 events with 1.5 expected background: 
\begin{equation}
 \sigma(e^+e^- \to \bar{d} X) < 0.031\;{\rm pb, \;at\;\sqrt{s} = 10.5\; GeV}.  
\end{equation}
Given that the continuum hadronic cross-section at $\sqrt{s} = 10.5$ GeV 
exceeds 3000 pb, we see that fewer than 1 in $10^5$ $q\bar{q}$ 
hadronizations results in anti-deuteron production, 
noticeably less than for $ggg,\,gg\gamma$ hadronization.  

\section{Conclusions}

Using CLEO data, we have studied anti-deuteron production 
from $\Upsilon(nS)$ resonance decays and the nearby continuum.  
$\Upsilon(1S)$ and $\Upsilon(2S)$ production rates are 
presented separately for the first time and combined to 
limit anti-deuteron production from $\chi_{bJ}(1P)$ states.  
First limits on production from the $\Upsilon(4S)$ and an 
improved continuum production limit are given.  

The results confirm a small but significant rate from hadronization 
of $\Upsilon(nS) \to ggg,\,gg\gamma$ decays, for $n=1,2$.  
However, no significant production from $q\bar{q}$ hadronization is 
observed; our $q\bar{q}$ limit is more than three times smaller 
than the observed rate from $ggg$ hadronization.  
Thus, the results indicate that anti-deuteron production 
is enhanced in $ggg,\,gg\gamma$ hadronization 
relative to $q\bar{q}$.  

We observe that baryon number conservation is accomplished with 
approximately equal amounts of accompanying $pp$, $pn$, $np$, $nn$.  
We also found three $d \bar d$ events; 
it is not immediately clear if double coalescence from initial 
baryons and anti-baryons can accommodate this rate, or if this 
is evidence for a more primary sort of (anti-)deuteron production 
in the hadronization process.


\section{Acknowledgments}

We gratefully acknowledge the effort of the CESR staff 
in providing us with excellent luminosity and running conditions.
A.~Ryd thanks the A.P.~Sloan Foundation.
This work was supported by the National Science Foundation
and the U.S. Department of Energy.

\end{document}